\documentstyle[]{article}
\begin{document}
\title{\large\bf Decoupling of Massive Right-handed Neutrinos}
\author{Paramita Adhya\\{\small\it Presidency College, 
Calcutta 700073, India}\\[5mm]D. Rai Chaudhuri\\{\small\it Presidency
College, Calcutta 700073, India}}
\date{}
\maketitle
\begin{abstract}{\small We investigate the effect of B+L$-$violating
 anomalous generation of massive right-handed neutrinos on
their decoupling, when the right-handed neutrino mass is considerably
greater than the right-handed gauge boson masses.  Considering normal
annihilation channels, the Lee-Weinberg type of calculation, in this
case, gives an {\it upper} bound of about 700 Gev, which casts doubt
on the existence of such a right-handed neutrino mass greater than
right-handed gauge boson masses. We examine the possibility that a
consideration of anomalous effects related to the $SU(2)_R$ gauge
group may turn this into a {\it lower} bound $\sim 10^2$ Tev.
\medskip

\noindent PACS number(s) : 18.80.-k, 14.60.St, 11.10.Wx. }
\end{abstract}

\begin{center}{\bf I. INTRODUCTION}\end{center}

Neutrino oscillation interpretation of recent observations of solar
and atmospheric neutrino fluxes, although presenting some
inconsistencies, may be taken to strengthen the idea of non-zero
neutrino masses. In this situation, in addition to the standard model
left-handed neutrinos, the existence and masses of right-handed
neutrinos assume topical interest.

The contribution of massive neutrinos to the mass-density of the
universe allows the setting of a lower bound to such a neutrino mass
from the usual cosmological constraints on the age and mass-density
of the universe [1,2,3]. The standard calculations consider a
neutrino mass less than gauge boson masses. In the present paper,
working in a L-R symmetric extension of the standard model [4,5,6],
we investigate how the nature of the bound is altered when the
right-handed neutrino mass is greater than gauge boson masses.

In these L-R symmetric models, the breaking of $SU(2)_R$ gauge
symmetry is  associated with a critical temperature. This may,
typically, be of the order of 1-10 Tev [7,8,9], and right- handed
electron neutrino masses $\approx$ 10 Tev have been considered,
yielding a left-handed electron neutrino mass $\approx 10^{-10}$ Gev,
by a see-saw mechanism [9]. Now, B+L is not conserved in standard
electro-weak theory due to an anomaly involving the SU(2) gauge group
[10], and, at temperatures $\ge$ 1 Tev, B+L$-$violating transitions
occur classically, via thermal fluctuations, at rates higher than the
expansion rate of the universe [11]. So, we may expect that similar
anomalous generation of right-handed neutrinos (in addition to the
left-handed ones), via the L-R symmetric gauge group, may become
important near the $SU(2)_R-$breaking critical temperature.

Although there is still a lot of fluidity in the matter, particle
physics and cosmological bounds usually suggest right-handed $Z_R$
and $W_R$ boson masses with values $\ge$ 0.5 TeV and 1.6-3.2 TeV,
respectively [12]. If, now, right-handed neutrinos of mass $\ge$ 10
TeV come under consideration in the literature, then it becomes
necessary to investigate whether anomalous effects can, indeed,
modify significantly the decoupling of right-handed neutrinos with
mass greater than right-handed gauge boson masses.

The plan of the paper is as follows. Section I is the Introduction.
In Section II, the L-R symmetric model is used to evaluate the
reduction rate of the right-handed neutrinos, in a standard
Lee-Weinberg type of calculation, and to observe how the cosmological
bound on their mass becomes an upper one, when this mass is greater
than the right-handed gauge boson masses.  In Section III, the
anomalous rate of reduction of right-handed neutrinos is related to
the general anomalous rate of B+L$-$violating transitions, and the
qualitative effect of the anomalous rate on the previously obtained
mass bound is estimated, assuming a generic form for the
B+L$-$violating rate arising from the anomaly involving the $SU(2)_R$
gauge group. In Section IV, the influence of these anomalous effects
on the mass bound is studied numerically, using numbers obtained by a
simple extrapolation from the $SU(2)_L$ result.\bigskip

\noindent\bf II.DECOUPLING WITHOUT ANOMALOUS EFFECTS\\A.Boltzmann 
equation for processes $N\bar{N}\to F\bar{F}$.\rm
\\\ We wish to set up a Bolzmann equation for the
number density of right-handed neutrinos, and, from a calculation of
the asymptotic number density, estimate the contribution of these
neutrinos to the mass-density of the universe, and, hence, set bounds
to the right-handed neutrino mass [1,3].

We will simplify matters by neglecting the decay of right-handed
neutrinos. Then, the normal electro-weak process chiefly responsible
for the reduction of right-handed neutrinos may be written as
$N\bar{N}\to F\bar{F}$, where F is a quark or a lepton, lighter than
N.

We are interested in investigating the situation when the right-
handed  neutrino mass is considerably greater than the right-handed
gauge boson masses.

To calculate the rate of reduction of right-handed neutrinos we
consider the L-R symmetric model [5,6]. This model has pairs of
fermion doublets $f'$  belonging to different representations of
SU(2)$_L\times$SU(2)$_R\times$U(1)$_{B-L}$, like\[
\left(\begin{array}{c}\nu_L\\ e_L\end{array}\right)\,
(\frac{1}{2},0,-1),\;
\left(\begin{array}{c}\nu_R\\
e_R\end{array}\right)\,(0,\frac{1}{2},-1);\] \[
\left(\begin{array}{c}u_L\\d_L\end{array}\right)\,
(\frac{1}{2},0,\frac{1}{3}),\;
\left(\begin{array}{c}u_R\\
d_R\end{array}\right)\,(0,\frac{1}{2},\frac{1}{3}).\] The numbers
refer to the quantum numbers $T_{3L},T_{3R},B-L$, respectively. We
will also write $\nu_R\equiv N,\nu_L\equiv\nu$.

The fermion gauge-boson interaction Lagrangian is
\[
L_{int}=g(\bar{f'}\gamma{_\mu} P_L\vec{T_L}f'.\vec{W_L{^\mu}} +\bar
{f'}\gamma{_\mu} P_R\vec{T_R}f'.\vec{W_R^{\mu}}
+\frac{1}{2}g'\bar{f'}\gamma_{\mu}(B-L)f'B^{\mu},
\] where $
P_{L,R}=\frac{1}{2}(1\pm\gamma_5),\,\vec{T}$ is the isospin operator
and $\vec{W{^\mu}},B{^\mu}$ are the gauge bosons.  The neutral
currents are set out in the basis\[
A{^\mu}=sin\theta(W{^\mu}_{3L}+W{^\mu}_{3R})+\sqrt[]
{cos2\theta}\,B{^\mu},\] \[
\hspace{-3mm}Z{^\mu}=cos\theta\,
W{^\mu}_{3L}-sin\theta tan\theta\,
W{^\mu}_{3R}-tan\theta\sqrt[]{cos2\theta}\,B{^\mu},\]
\[  Z'{^\mu}=\frac{\displaystyle
\sqrt[]{cos2\theta}}{\displaystyle
cos\theta}\,W{^\mu}_{3R}-tan\theta\, B{^\mu},\]where $\theta$ is the
Weinberg angle.\\Neglecting Z-Z$'$ mixing, one gets the Z$'$ neutral
current Lagrangian [12]\begin{eqnarray*}\lefteqn
{L_{NC}^{Z'}=\frac{\displaystyle g}{\displaystyle
cos\theta\sqrt[]{cos2\theta}}(sin^2\theta\sum_{f'}\bar{f'}
\gamma{^\mu}[P_LT_{3L}-Qsin^2\theta]f'+}\\ & & cos^2\theta
\sum_{f'}\bar f'\gamma{^\mu}[P_RT_{3R}-Qsin^2\theta]f')
Z'{_\mu}.\quad (1)\end{eqnarray*} Q is the charge operator.
\\The charged current Lagrangian consists of terms of the form
\[L{_{CC \nu e}}=\frac{\displaystyle g}{\displaystyle\sqrt[]{2}}(\bar
{\nu}\gamma{_\mu} e_L W{^{\mu}}_L{+h.c.})+\frac{\displaystyle g}
{\displaystyle \sqrt[]{2}}(\bar{N}\gamma{_\mu}e_RW{^{\mu}}_R{+h.c.}).
\qquad (2)\] Assuming CP-symmetry, and equilibrium conditions for all
relevant particles except the N neutrinos, the rate of reduction of N
neutrinos per unit volume can be obtained from the Boltzmann
collision integral for the processes $N\bar{N}\to F\bar{F}$ [13,14]:
\begin{eqnarray*}\lefteqn{\hspace{-2.5cm}\Gamma_a=\sum_F\int d\pi_Nd
\pi_{\bar{N}}d\pi_Fd\pi_{\bar{F}} (2\pi)^4\delta^4(p_N+p_{\bar{N}}-
p_F-p_{\bar{F}})}\\& & |{\cal M_F}|^2(f_Nf_{\bar
{N}}-f_{Neq}f_{\bar{N}eq}).
\quad(3)\end{eqnarray*}
Here, $ f$ is the phase space distribution function and $f_{eq}$ is
its equilibrium  value.  $|{\cal M_F}|^2$ is the spin averaged matrix
element squared, with proper symmetry factor, for the process
$N\bar{N}\to F\bar{F}$, assumed, by CP-symmetry, to be the same as
that for the process $F\bar{F}\to N\bar{N}$. The measure
$d\pi_i=g_i\, d^3p/((2\pi)^3 2E),\;g_i$ being the degeneracy number.
We assume that there is no significant Fermi degeneracy, so that
$1-f\approx 1.$

 Because CP-symmetry has been assumed, we further assume that there
is no N or $\bar{N}$ excess, and we can set  $n=\bar{n}$, as well as
$\mu_N=0=\mu_{\bar{N}}$. We can, then, take
\[f_{Neq}= e^{-{E_N}/T}.\]

The summation is over quarks and leptons lighter than N.  Let us take
$\nu$ and N to be electron neutrinos. We assume right-handed
neutrinos of the other two generations to be much more massive than
the N neutrinos, so that they are not relevant here.

It is usual to introduce the thermal average of the annihilation
cross-section times relative velocity

\begin{eqnarray*}\lefteqn{\hspace{-36mm}<\sigma|v|>
=\frac{\displaystyle 1}{\displaystyle n_{eq}^2}\sum_F
\int d\pi_N d\pi_{\bar{N}}d\pi_F d\pi_{\bar{F}}\,
(2\pi)^4
\delta^4(p_N+p_{\bar{N}}-p_F-p_{\bar{F}})}
\\ & & |{\cal M_F}|^2e^{-E_N/T}e^{-E_{\bar{N}}/T}, 
\end{eqnarray*}and write (3) in the form [1]\[\Gamma_a=<\sigma|v|>
(n^2-n_{eq}^2),\]
\noindent where n is the number density of the N neutrinos and
 $ n_{eq} $ is its equilibrium value.

Then the Boltzmann equation for the reduction of N neutrinos by these
processes, in a universe expanding with $\dot{R}/R= H,$ becomes [14]
\[\frac{\displaystyle dn}{\displaystyle
dt}+3Hn=-<\sigma|v|>(n^2-n_{eq}^2).\qquad(4)\]

\noindent\bf B.Calculation of $<\sigma|v|>$ \bf from L-R 
symmetric model.

\noindent\rm The Feynman diagram for $ N\bar{N}\stackrel
{Z'}\to F\bar{F}$, from (1), is given in Fig. 1.  For the
$N\bar{N}\stackrel{W}\to e\bar{e}$ amplitude, we get, using (2), an
additional contribution from the diagram shown in Fig. 2.
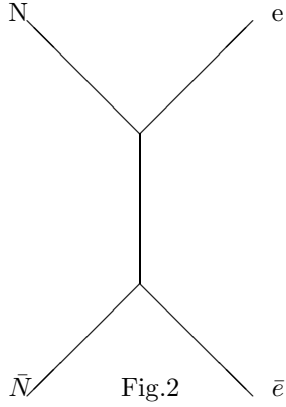
\setlength{\unitlength}{0.5cm}
\begin{picture}(12,8)\put(4,4){\line(1,0){4}}\put(4,4)
{\line(-1,1){3}}\put(4,4){\line(-1,-1){3}}\put(8,4)
{\line(1,1){3}}\put(8,4){\line(1,-1){3}}\put(1,7){N}
\put(0.5,1){$\bar{N}$}\put(11,7){F}\put(11,1){$\bar{F}$}
\put(2,6){k}\put(2.5,2){$\bar{k}$}\put(10.5,6){p}
\put(10,2){$\bar{p}$}\put(5.5,1){Fig.1}\end{picture}\\ 
\begin{picture}(8,12)\put(4,4){\line(-1,-1){3}}
\put(4,4){\line(1,-1){3}}\put(4,8){\line(1,1){3}}
\put(4,8){\line(-1,1){3}}\put(0.5,1){$\bar{N}$}
\put(0.5,11){N}\put(7.5,11){e}\put(7.5,1){$\bar{e}$}
\put(3.5,1){Fig.2}\put(4,4){\line(0,1){4}}\end{picture}
\medskip

We will work at temperatures $T<T_{CR}$, the critical temperature
corresponding to the breaking of $SU(2)_L\times SU(2)_R\times
U(1)_{B-L}$ to $SU(2)_L\times U(1)_Y.$ We are going to consider
N-type neutrinos with a mass M, which is at least an order or two of
magnitude larger than $M_{Z'} (M_{Z'}\ge 0.5$ Tev [12]). At this
energy scale, we will approximate all quark and the e,$\mu,\tau,\nu$
masses to zero (mass of top$\approx$175 Gev).

Next, we assume that $\nu$ and N have Majorana mass eigenstates
[15]\[\qquad\chi=\nu+\nu^c,\qquad\omega=N+N^c,\]
\noindent where the superscript "c" refers to the charge conjugate
field.

It is usual to consider a bidoublet and two triplet Higgs particles
to generate Majorana states [6]. In this paper, however, we do not go
into the details of any specific model of the Higgs sector.
 While evaluating the matrix element, we have considered N to be
purely Majorana, i.e. we have neglected the contribution of the
vector current and doubled that of the axial current by replacing
$(1+\gamma_5)$ with $ 2\gamma_5$ [12].

The spin-averaged matrix element squared, with symmetry factor
$\frac{\displaystyle 1}{\displaystyle 2!}$,  for the first diagram
gives, from (1),
\begin{eqnarray*}\lefteqn{|{\cal M_F}|^2=\frac{1}{2}
[g^4/(2cos^{2}2\theta)] (C_{VF}^2+C_{AF}^2)}\\ & &
[(p.k)(\bar{p}.\bar{k})+
(\bar{p}.k)(p.\bar{k})-M^2p.\bar{p}]\frac{\displaystyle 1}
{\displaystyle (q^2-M_{Z'}^2)^2},
\end{eqnarray*} where,
\[C_{VF}=T_3-2Q\,sin^2\theta,\quad
C_{AF}=T_3\,cos2\theta.\] Now,\begin{eqnarray*}
\lefteqn{q^2=(k+\bar{k})^2=s= 4E_{CM}^2 }\\ & &
=4(M^2+k_{CM}^2) \\ & & >> M_{Z'}^2,
\end{eqnarray*} where $(E_{CM},{\bf k}_{CM})$ is
the 4-vector k in the CM frame.  So, we approximate
$1/(q^2-M_{Z'}^2)^2$ by $1/q^4$.

We calculate $<\sigma|v|>$ in two steps.\\ First, we calculate
\[I_F=\int d\pi_Fd\pi_{\bar{F}}(2\pi)^4 
\delta^4(k+\bar{k}-p-\bar{p})|{\cal M_F}|^2 \] in the 
CM frame.The result is\[I_F=[g^4/(64\pi
cos{^2}2\theta)](C_{VF}^2+C_{AF}^2)(2/3)\beta^2,\] where $\beta$ is
the relative velocity = $2|\bf{k}_ {\rm{CM}}|/\rm{E_{CM}}$. This
"p-wave" term is a signature of Majorana neutrino annihilation.

In the Lee-Weinberg type of decoupling calculation, the N neutrinos
may be considered to be non-relativistic, as the relevant
temperatures are of the order of M. Then, in the comoving "lab"
frame, where $\bf\bar{k}$ \rm makes an angle $\alpha$ with $\bf{k}\rm
$,\[I_F=[g^4/(64\pi cos{^2} 2\theta)](C_{VF}^2+C_{AF}^2)
(2/3)(\bf{k}\rm^2+\bf{\bar{k}}\rm^2-2|\bf{k}||\bar{k}|
\rm{cos}\alpha)/M^2.\quad (5)\] In the second step we do the
 thermal averaging. Then,
\[<\sigma|v|>_F=\frac{\displaystyle \int d\pi_N 
e^{-E_N/T}\int d\pi_{\bar{N}}e^{-E_{\bar{N}}/T} I_F}{\displaystyle
\int\frac{\displaystyle g_Nd^3k} {\displaystyle (2\pi)^3}
e^{-E/T}\int\frac{\displaystyle g_{\bar{N}}d^3\bar{k}}{\displaystyle
(2\pi)^3} e^{-E_{\bar{N}}/T}}.\] Calculation, in the non-relativistic
approximation for the N neutrinos, gives\[<\sigma|v|>_F=
\frac{\displaystyle g^4}{\displaystyle (64\pi cos^2{2}\theta)}
(C_{VF}^2+C_{AF}^2)\frac {\displaystyle 1}{\displaystyle
M^2}\frac{\displaystyle T} {\displaystyle M}.\] In the case of
$N\bar{N}\to e\bar{e}$, the effect of the extra diagram can be taken
into consideration by the usual Fierz rearrangement, which gives, in
this case,\begin{eqnarray*} C_{Ve}/cos2\theta\to
(C_{Ve}/cos2\theta)\,+1,\\ C_{Ae}/cos2\theta\to
(C_{Ae}/cos2\theta)\,+1.
\end{eqnarray*} Finally, we get
\[ <\sigma|v|> =\frac{\displaystyle g^4}{\displaystyle 
(64\pi cos^22\theta)}\frac{\displaystyle 1}{\displaystyle
M^2}\frac{\displaystyle T}{\displaystyle M}
\sum_F(C_{VF}^2+ C_{AF}^2).\quad (6)\]
So, effectively, $<\sigma|v|>\sim\frac{\displaystyle 1}
{\displaystyle M^2}$, as $T\sim M.$\medskip

\bf C.Mass Bound for Right-handed Neutrinos.\rm

Introducing x=M/T and Y=n/s, where s is the entropy density, (4)
becomes\[(1.66g^{*\frac{1}{2}}/x^4)(M^5/M_{Pl})
\frac{\displaystyle 2\pi^2}{\displaystyle 45}
g^*_s\frac{\displaystyle dy}{\displaystyle dx}= -(\frac{\displaystyle
2\pi^2}{\displaystyle 45} g^*_s)^2\frac{\displaystyle
M^6}{\displaystyle x^6}<\sigma|v|>(Y^2-Y_{eq}^2).\quad(7)\] or,\[
\frac{\displaystyle dY}{\displaystyle dx}=
-0.26g^{*\frac{1}{2}}<\sigma|v|>(MM_{Pl}/x^2) (Y^2-Y_{eq}^2).\quad
(8)\] We take $M_{Pl}=1.22
\times 10^{19}$ Gev, and $g^*\approx g_s^*\approx 
100$ just below the critical temperature, considering
$N,W_R,Z'$(and $N_{\mu},N_{\tau})$ to be massive (we have not
counted Higgs degrees of freedom).
\\Summing over all quarks and leptons, except the three right-
handed neutrinos, we get, on calculation,\[
\sum_F(C_{VF}^2+C_{AF}^2)=3.28\quad\mbox
{(taking}\quad sin^2\theta=0.23).\] Taking g=0.65,
$<\sigma|v|>=0.01/(M^2x).$\\ For massive Majorana neutrinos, we get,
in the non-relativistic approximation [14],\[Y_{eq}=2.89\times
10^{-3}x^{\frac{3}{2}} e^{-x}.\qquad(9)\] From (8),
\[\frac{\displaystyle dY} {\displaystyle
dx}=-(3.16\times 10^{17}/Mx^3) (Y^2-Y_{eq}^2).\qquad (10)\]

 We write  $\Delta=Y-Y_{eq}$.\\Then, before decoupling, Y$\approx
Y_{eq}$, and $\Delta'\sim 0$, giving\[ \Delta
\cong - Mx^3Y'_{eq}/[3.16\times 10^{17}(2Y_{eq}+
\Delta)]\] Now, we put $\quad \Delta=cY_{eq}\quad$ at 
decoupling, where c$\sim 1$. According to the general numerical
analysis of this type of decoupling [14],$\; c(c+2)=2\quad $when
$<\sigma|v|>\sim T.$

At decoupling, when $x=x_d>>1, Y'_{eq}\approx -Y_{eq}, $ and
\[ \Delta(x_d)\cong cY_{eq}(x_d) =Mx_d^3/[3.16\times 10^{17}(c+2)].
\qquad(11)\] This leads to \[ x_d\cong 35.14-lnM-1.5\,ln(35.14-lnM).
\qquad  (12)\]

After decoupling, $\quad Y>>Y_{eq}\quad $and $
\Delta\approx Y$. \\From (10), we get \[\Delta'=
-3.16\times 10^{17}\Delta^2/(Mx^3),\] which gives, on integration, at
$t\to\infty, ${}\[\Delta_{\infty}=Y_{\infty}=2Mx_d^2 /(3.16\times
10^{17}),\] assuming,$\quad Y(x_d)>> Y_{\infty}.$

We will take as our cosmological bound [14]:\[
\Omega_Nh^2<1, \; \mbox{where}\;
\Omega_N=\rho_N/\rho_c=Ms_0Y_{\infty} /\rho_c.\] Here, it is assumed
that  h$>0.4$. \\Taking $s_0=2970\, cm^{-3}\quad$and$\quad \rho_c=h^2
1.88\times 10^{-29}gcm^{-3}$, we get
\begin{eqnarray*}\lefteqn{\Omega_Nh^2} \\ & & =
2.80851\times 10^8 MY_{\infty}\qquad\qquad (13)
\\ & & =3.62\times 10^{-9}M^2x_d^2,\qquad\qquad
 (13a)\end{eqnarray*} where M is to be taken in Gev.\\ At the bound,
\[3.62\times 10^{-9} \,M^2x_d^2=1.\] Solving this equation 
and (12) simultaneously, using simple numerical methods, we get
\[x_d=23.55,\qquad M=706\: Gev.\]Now, if we omit ln M in the third
term on the R.H.S. of (12), we get, approximately, \[
x_d=29.80-lnM.\qquad\qquad (14)\] If we make this approximation, the
error in $x_d$ is less than 5 percent, even if M is as large as
$10^6$ Gev.  Using (14), we get \[
d(\Omega_Nh^2)/dM=3.62\times10^{-9} \times
2M(29.80-lnM)(28.80-lnM),\] which is positive for all practical
purposes.\\This means that $\Omega_Nh^2<1$ fixes an \it upper \rm
bound for M.

This can be seen transparently if we work in the approximation
\[Y_{\infty}\approx Y(x_d)\approx 2Y_{eq}(x_d),\]
\[ \hspace{-3mm}  \mbox{taking c}\approx 1 \;
\mbox{in}\;(11).\]Then,\[ \Omega_Nh^2\sim Mx_d^{
\frac{3}{2}}e^{-x_d}.\] (14) shows that $\Omega_Nh^2
\sim M^2(29.80-ln\,M)^{\frac{3}{2}}$, and, so, 
as M increases, $\Omega_Nh^2$ increases, for all practical values of
M.

This conclusion can be verified, numerically, by giving M different
values in (12) and substituting the resulting $x_d$ in (13a). The
results are shown in Table 1.

In the usual Lee-Weinberg case, with $M<<$ gauge boson masses, one
gets a \it lower \rm bound because $<\sigma|v|>
\sim M^2$, which leads to $x_d\sim 3\,ln\,
M+constant$, and $\Omega_Nh^2\sim
\frac{\displaystyle 1}{\displaystyle M^2}$. With
$M>>$ gauge boson masses, $<\sigma|v|>\sim
\frac{\displaystyle 1}{\displaystyle M^2}$ , and
this makes the difference.

\begin{table}\begin{center} {\bf Table 1\\Mass Bound (no 
anomalous effects)}\\[1ex]\begin{tabular}[]{cc} M(Gev) &
$\Omega_Nh^2$\\10,000 & 160\\5,000 & 42.6\\1,000 & 1.95\\750 &
1.12\\706 & 1.00\\ 500 & 0.516\\250 & 0.136\end{tabular}\end{center}
\end{table}

$M < 706$ Gev is, in effect, incompatible with our assumption of $M
>$ right-handed gauge boson masses, because, as we have remarked
earlier, particle physics and cosmological bounds suggest
right-handed gauge boson masses $\sim 0.5-1$ Tev or more. We have to
conclude that the assignment of any realistic mass, greater than
right-handed gauge boson masses, to the right-handed neutrinos will
violate the cosmological bound $\Omega_N h^2<1$.

However, we have considered only the normal elecrtroweak process
$N\bar{N}\to F\bar{F}$. In the next section, we consider, in
addition, anomalous processes.\bigskip

\noindent\bf III. INTRODUCTION OF ANOMALOUS EFFECTS.

\noindent A. Anomalous generation of right-handed neutrinos.\rm

For the standard model, a classical, unstable, time-independent
solution of the equations of motion has been identified [16,17].
This sphaleron solution corresponds to the barrier between vacua with
different topological numbers.  A sphaleron-mediated transition over
the barrier leads to a fermion-number violating transition with
$|\Delta L|=3,|\Delta B|=3$, of the type\[
|\,W^{cl}_{\mu}\,\alpha\,>\,{\longrightarrow}\,
|\,W^{'cl}_{\mu}\,\alpha '\,>,\] where $\alpha,
\alpha '$ are fermion states, differing by\\ $|\Delta L|=3,|
\Delta B|=3$, and $W^{cl}_{\mu},W^{'cl}_{\mu}$ 
are the initial and final SU(2) gauge boson configurations, which are
essentially classical.  (We are neglecting the small effect of the
U(1) part [17].)

All colours and families of quarks and leptons will be generated
equally, but, in any one transition, only one member per doublet will
be found. For the rest of this paper, we will neglect family mixing
and consider anomalous generation for a single (the lightest) family.
In this case, $|\Delta L|=1,|\Delta B|=1$.  $\alpha,\alpha '$ will be
restricted by the requirement that the sphaleron must be a colour
singlet,  SU(2) singlet and neutral mediator.  There are then two
relevant amplitudes, which we may write formally as \\ 
$<W^{cl}_{\mu}uudeW^{'cl}_{\mu}>$ and
$<W^{cl}_ {\mu}udd\nu W^{'cl}_{\mu}>$. \\All processes with these
amplitudes can occur. For example, $\alpha$ may be the vacuum and
$\alpha '$ may represent uude or udd$\nu$.

In the L-R symmetric model, we expect, on general grounds [16,11,9],
anomalous B+L generation above or just below $T_{CR}$, from both
$SU(2)_L$ and $SU(2)_R$ gauge boson field configurations with
non-vanishing topological charge.

 However, the actual construction of the sphaleron solution depends
on the details of the Higgs multiplet. The $SU(2)_L$ sphaleron [17]
was worked out with a complex doublet. In the L-R symmetric case, the
generation of Majorana masses at the higher energy scale results from
spontaneous symmetry-breaking associated with a $SU(2)_R$ triplet
scalar field ( in addition to a $SU(2)_L$ triplet and a bidoublet
which develop v.e.v. at the lower energy scale ) [6]. It has been
shown [9] that the topological condition necessary for the existence
of a sphaleron solution is fulfilled for a simplified model of
$SU(2)_R$ symmetry-breaking at the higher energy scale via a triplet
complex scalar field. But, the construction of an explicit solution
has proved very difficult.

In this situation, one has to assume [9,18] the occurrence of B+L
$-$violation via sphalerons for the $SU(2)_R$  gauge group, in
addition to B+L$-$violation for $SU(2)_L$ at the higher energy scale.
Neglecting mixing parameters between left-handed and right-handed
gauge bosons, we work with a highly simplified model in which the
$W^{\mu}_L$ give rise to anomalous generation of leptons and baryons
from left-handed doublets, and the $W^{\mu}_R$ from right-handed
doublets. In particular, the $W^{\mu}_R$ will generate, anomalously,
right-handed N neutrinos.

First, we want to relate the rate of production of the right-handed N
neutrinos per unit volume to the total rate $\Gamma$ of $
\Delta B=1,\Delta L=1$ anomalous transitions per unit volume. 
We divide $\Delta L=+1$ processes into four types (assuming distinct
flavour eigenstates for N and $\bar{N}$):\\ $l$: processes with a N
in the final state, e.g.\[ |W^{cl} _{\mu R}\,vac>\to |W^{'cl}_{\mu
R}\,uddN>,\]{} $\bar{l}$: processes with a $\bar{N}$ in the initial
state, e.g.\[ |W^{cl}_{\mu R}\,\bar{N}
\bar{u}\bar{d}\bar{d}>\to |W^{'cl}_{\mu R}\,
vac>,\]{}$m$: processes with an $e^-$ in the final state, e.g.\[
|W^{cl}_{\mu R}\,vac>\to|W^{'cl}_ {\mu R}\,uude^->,\]{} $\bar{m}$:
processes with an $e^+$ in the initial state, e.g.\[ |W^{cl}_{\mu
R}\,\bar{u}\bar{u}\bar{d}e^+>\to|W^{'cl}_ {\mu R}\,vac>.\]
Therefore,\[
\Gamma=\sum_l\Gamma_l+\sum_{\bar{l}}\Gamma_{\bar{l}}
+\sum_m\Gamma_m+\sum_{\bar{m}}\Gamma_{\bar{m}},\] where $\sum_i$ is a
sum over all processes of type $i$.

Each process has a rate which is determined in an essentially
classical way:  if the thermal fluctuation has sufficient energy to
cross the barrier, the process will occur. If $i\omega^-$ is the
frequency of the unstable (sphaleron) mode, a classical statistical
mechanics calculation gives [19,20]\[
\Gamma_i=(\omega^-/\pi)(Im{\cal
F}/T),\qquad\qquad (15)\] where ${\cal F}$ is the free energy. Also,
\[ (Im{\cal F}/T)\sim e^{-(V_0/T)},\qquad\qquad (16)
\] where $V_0$ is the barrier height.\\Because of this essentially
classical nature, the barrier-crossing rate,
at a given temperature, under equilibrium conditions, should 
 be of the same order in different channels. In other
words, we may expect that the rate of $\Delta L=1,\Delta B =1$
transitions, featuring one member of a lepton doublet, will be of 
the same order as the rate of such transitions, featuring the 
other member of the doublet. As a first approximation, we may
take\[\sum_l\Gamma_l+\sum_{\bar{l}}\Gamma_ {\bar{l}}\approx
\sum_m\Gamma_m+\sum_{\bar{m}}
\Gamma_{\bar{m}}\approx \frac{1}{2}\Gamma.\quad
 (17)\]
The approximation will be bad when the N neutrinos are way out of
equilibrium. In a decoupling study, however, one is interested in
finding out when the species just falls out of equilibrium.

Let us now interpret $l,\bar{l}$ formally as Boltzmann collisional
processes\[ l:  i + j +
.....\,\longrightarrow\quad N + a + b +
.....\;.\] \[ \bar{l}:\bar{N} + \bar{a} + \bar{b}
+  .....\,\longrightarrow\quad \bar{i} + \bar{j} + .....\; .\] CPT
ensures that for every process of type $l$, there is a process of
type $\bar{l}$ with the same matrix element ${\cal M}_l$. Then, we
can write, formally, [13,14]\begin{eqnarray*}\lefteqn{ \Gamma_l=\int
d\pi_Nd\pi_ad\pi_b...d\pi_id\pi_j...| {\cal M}_l|^2}\\ & & (2\pi)^4
\delta^4(p_N+p_a+p_b...-p_i-p_j..)\,
f_N^{eq}f_af_b...\;.\quad (18)\end{eqnarray*} We have again assumed
that all relevant species are in equilibrium except the right-handed
neutrinos, and that there is no significant Fermi degeneracy or Bose
condensation. Also,\begin{eqnarray*}
\lefteqn{ \Gamma_{\bar{l}}=\int d\pi_{\bar{N}}d\pi_{\bar{a}}d\pi_
{\bar{b}}...  d\pi_{\bar{i}}d\pi_{\bar{j}}...|{\cal M}_l|^2}
\\ & & (2\pi)^4 \delta^4(p_{\bar{N}}+p_{\bar{a}}+p_{\bar{b}}...
-p_{\bar{i}}-p_{\bar{j}}..)f_{\bar{N}}f_
{\bar{a}}f_{\bar{b}}...\:.\quad(19)\end{eqnarray*} In these formal
expressions, $|{\cal M}_l|^2$ is related to the classical probability
and is not to be interpreted perturbatively.

As we are interested here in decoupling and not in baryogenesis, we
will neglect  small CP-asymmetric effects and assume CP-symmetry.
Then, we can assume, as in Sec.II.A.,
\[ n={\bar{n}},\, n_{eq}=\bar{n}_{eq};\;\mbox{also,}\; f_a=
e^{-\frac{E_a}{T}},\,f_{\bar{a}}=e^{-
\frac{E_{\bar{a}}}{T}}\,\mbox{etc.}\quad (20)\] 
In this case, we can write, from (18) and (19),\[
\Gamma_l=I_l\,n_{eq},\quad\mbox{and} \quad
\Gamma_ {\bar{l}}=I_l\,n,\quad (21)\] where $I_l$
contains the result of the phase space integrations, apart from
$n_{eq}$ and $n$ [14,21]. ($I_l=I_{\bar{l}}$ due to (20).) It can be
interpreted as a thermally averaged width in mode
$l$. \\From (17) and (21), \[
\sum_l I_l(n+n_{eq})=\frac{1}{2}
\Gamma,\quad\mbox{and}\] \[ \sum_l\Gamma_l=
\frac{\displaystyle n_{eq}}{\displaystyle 2(n+n_{eq}
)}\:\Gamma.\quad(22)\] For $\Delta L=-1$ anomalous transitions, we
will get, similarly, a process $l'$ with a $\bar{N}$ in the final
state, and a process $\bar{l}'$ with a N in the initial state, and a
similar result
\[ \sum_{\bar{l}'}\Gamma_{\bar{l}'}=\frac{
\displaystyle n}{\displaystyle 2(n+n_{eq})}\:
\Gamma'.\quad(23)\] Now, we are neglecting baryon- and lepton-
number excess or deficit. In this approximation, we can set
$\mu_N=0$. We can, then, take [20,22]$\quad
\Gamma=\Gamma'$,  i.e. the rate of $\Delta L
=1$ transitions $\approx$ the rate of $\Delta L=-1$ transitions.

Hence, the net rate of reduction of N neutrinos by anomalous
processes, per unit volume, $\Gamma_N$ = rate of such processes, per
unit volume, with a N in the initial state $-$ rate of such
processes, per unit volume, with a N in the final state\[
=\sum_{\bar{l}'}\Gamma_{\bar{l}'}-
\sum_l\Gamma_l.\]

We finally get\[ \Gamma_N=\frac{\displaystyle n-n_{eq}}{\displaystyle
2(n+n_{eq})}\,\Gamma.\qquad (24)\] As expected, the anomalous rate
vanishes if the N neutrinos are in equilibrium.

Since, we have Majorana mass states, lepton number violating
processes with $\Delta L=2$ can arise from L-violating terms in the
Lagrangian. A variety of such processes has been considered in the
literature [9,18,23]. These theories are, broadly, of two types.

In one type of theory, a massive right-handed Majorana neutrino is
allowed to decay. As we are neglecting decay, we have not considered
such theories.

In the other type of theory, the L-violating term is, effectively, of
the type $(m_{\nu}/v^2)\nu\nu HH$, where H is a Higgs scalar,
$m_{\nu}$ is the $\nu$ mass, and $v$ is the Higgs v.e.v. Usually, in
these theories, one takes, $v^2/m_{\nu}$ = M.  If we consider a
similar term (1/M)NNHH, we will find a cross-section $\sim (1/M^2)$.
So, such terms will not give processes significantly faster than
$N\bar{N}\to F\bar{F}$, with $M>>$gauge boson masses, and may be left
out in an exercise where the emphasis is on qualitative features of
the decoupling.

Assuming, therefore, CP-symmetry and equilibrium conditions for all
relevant particles except right-handed neutrinos, the rate of
reduction, per unit volume, of the N neutrinos can be written in the
form of a Boltzmann equation\[
\frac{
\displaystyle dn}{\displaystyle dt}+3Hn=-\frac{
\displaystyle (n-n_{eq})}{\displaystyle 
2(n+n_{eq})}\:\Gamma_R -<\sigma|v|> (n^2-n_{eq}^2).\qquad(25)\] This
is our basic equation.\\We have put $\Gamma=\Gamma_R$ to indicate
that we are considering the anomalous rate for right-handed
neutrinos.\medskip

\noindent \bf B.Anomalous effects in decoupling.\rm

For $T<T_{CL},$ the critical temperature for the spontaneous
breakdown of $SU(2)_L\times U(1)_Y,\;\Gamma =\Gamma_L$ has been
calculated [20,24]. For the right-handed case, with $T<T_{CR}$, the
complication of the Higgs sector has obstructed a calculation of
$\Gamma_R$.  However, the very general considerations mentioned in
Sec.III.A. imply that $\Gamma
\sim e^{-V_0/T}$.

$V_0$ can be estimated heuristically, as follows [25,26]. If we
assume a sphaleron solution with energy $E_{sp}$, we can put
$V_0=E_{sp}$. But $E_{sp}$ arises from a classical solution, i.e.
from a limit where many quanta are involved. We can take the energy
per quantum $\sim M_W$, and the average number of quanta $\sim
(1/\alpha_W)$, where  $\alpha_W=\frac {\displaystyle
g^2}{\displaystyle 4\pi}$. Then, $E_{sp}\sim
\frac{\displaystyle M_W}{
\displaystyle \alpha_W}$  is a very general 
estimate, which should hold for the right-handed case also, with
$M_W=M_{WR}$.  So, we can write, on general grounds,\[
\Gamma_R=\tilde{R}(M_{WR},T)e^{-B\,M_{WR}
/(\alpha_WT)},\qquad(26)\] where $\tilde{R} ,B$ depend on the precise
form of the symmetry-breaking.  However, we can say that $\tilde{R}$
will have dimension$\sim$(mass)$^4$, and B will be dimensionless and
of order 1. Also, the whole idea of separating out the exponential is
to isolate a prefactor $\tilde{R}$ which can be assumed to vary more
slowly (in the left-handed case, the prefactor varies as powers of
the arguments [20]).

Introducing x and Y, the Boltzmann equation (25) becomes, from (7),
(10), and (26),\begin{eqnarray*}\lefteqn {\hspace{-28mm}
\frac{\displaystyle dY} {\displaystyle dx}=
-\frac{\displaystyle 1} {\displaystyle
(1.66g^{*{\frac{1}{2}}}/x^4)(M^5/M_{Pl})
\frac{2\pi^2}{45}g^*_s}\:\frac{\displaystyle
 Y-Y_{eq}}{\displaystyle 2(Y+Y_{eq})}
\tilde{R}'(M,a,x)\,e^{-Bx/(\alpha_W\,a)}}\\ & &
-\frac{\displaystyle 3.16\times10^{17}} {\displaystyle
Mx^3}(Y^2-Y_{eq}^2),\quad(27)
\end{eqnarray*} writing $a=(M/M_{WR})$. We are
 considering a$>$1.\\Compactly, we can write\[
\frac{\displaystyle dY}{\displaystyle dx}=-\,R(M,a,x)e^{-Bx/
(\alpha_Wa)}\frac{
\displaystyle Y-Y_{eq}}{\displaystyle 2(Y+
Y_{eq})} -\,\frac{\displaystyle 3.16\times 10^{17}}{\displaystyle
Mx^3}(Y^2-Y_{eq}^2).
\quad(28),\]where $\tilde{R}'$ and R are obtainable
 from $\tilde{R}$.

As "a" (the quotes are to avoid confusion with the singular a)
increases, the first term gains importance because of the
exponential. Suppose "a" has a value for which the first term is
predominant near decoupling. Let us characterise decoupling by the
simple criterion\[
\Delta(x_d)=Y_{eq}(x_d).\qquad\qquad(29)\] This
is equivalent to taking c=1 in (11).\\ Before decoupling, $Y\approx
Y_{eq},\Delta'
\sim 0,$ and\[ Y_{eq}'=-\,Re^{-Bx/(\alpha_Wa)}
\frac{\displaystyle \Delta}{\displaystyle 2(2Y_{eq}
+\Delta)}.\qquad(30)\] At decoupling, when $Y_{eq}' =-Y_{eq}$, (30)
gives\[{} [6Y_{eq}(x_D)/R] e^{Bx_d/(\alpha_Wa)}=1,\] and, using the
value of $Y_{eq}$ from (9),\[
e^{[\frac{B}{\alpha_Wa}\:-1]x_d}\,G(M,a,x_d) =1,\] where the
prefactor G can be assumed to have a slower variation with $x_d$ and
"a" than the exponential, because R is a prefactor for which this has
been assumed.  Then, assuming the exponential to dominate, we expect,
approximately, \[(\frac{\displaystyle B}{\displaystyle
\alpha_Wa}\,-\,1)x_d=\tilde{B}x_d\approx
constant.
\quad (31)\]\hspace{3mm} Now, the sphaleron decay 
will produce a N neutrino only if the kinematic constraint $E_{sp}>M$
is satisfied. As $E_{sp}=BM_{WR}/
\alpha_W$, so, $B/\alpha_W\,>a$ gives an upper
 limit $a'$ on "a" for anomalous effects to occur. For $a<a',\;\tilde
{B}>0,$ and, if "a" is increased, $\tilde{B}$ decreases, so that,
$x_d$ increases.

We approximate $Y_{\infty}$ by $Y(x_d)$, so that, from (29),\\
$Y_{\infty}\approx 2Y_{eq}(x_d)$.  Then, we get, from (13) and (9),\[
\Omega_Nh^2= 1.62332\times 10^6\,x_d^{\frac{3}{2}}\,
e^{-x_d}\,aM_{WR}.\qquad(32)\] Since, from (31),
$x_d\sim\,\frac{a}{a'}/(1-\frac{a}{a'})$, the exponential will
dominate, and we can expect that, as "a" increases, $\Omega_Nh^2$
will decrease. This means that $\Omega_Nh^2<1$ will give a lower
bound on "a", and, hence, on M, for a given $M_{WR}$, for a$<a'$.

If we can actually find values of the parameter "a"= $M/M_{WR}$,
within the range $1<a<a'$, for which the anomalous term in (28)
predominates, there will not be any hindrance from the Lee-Weinberg
type of cosmological bound to right-handed neutrinos having masses
greater than right-handed gauge boson masses.

 So, we find that anomalous reduction of right-handed neutrino number
may have important effects on the decoupling of such neutrinos.

Whether these formal expectations will be borne out depends on the
actual numbers in $\Gamma_R$.  Extrapolating the known result for
$\Gamma_L$ to the right-handed case, keeping wide leeway, we will
find that numerical results give cause for optimism.
\bigskip

\noindent \bf IV. NUMERICAL RESULTS\rm

We will take the $\Gamma_L$ given in Reference [20]. \[
\Gamma_L=\frac{\displaystyle (1.4\times 10^6)\, M_W^7}{\displaystyle
g^6\,T^3}\,exp[-\frac {\displaystyle 16\pi M_W}{\displaystyle
g^2\,T}].
\quad (33)\] Here, the unstable mode $\omega^-$ 
is taken $\approx M_W$, and $E_{sp}=2\,(M_W/\alpha_W)
\,\bar{E}.\quad \bar{E}$ is a number which depends on 
$(\lambda/g^2)$:\,$ 1.56<\bar{E}<2.72$ for $0<\lambda<\infty
(\lambda$ is the
4-Higgs self-coupling constant). We
take $\bar{E}=2.
\quad M_W$ is temperature dependent.\[ M_W=M_W(0)[1-(T/T_C)^2]^
{\frac{1}{2}},
\qquad (34)\] and $T_C=3.8\,M_W(0)$ [20]. 
There is an overall constant $\kappa\sim 1$ [20,24].  We take
$\kappa=1$.

This expression is valid for $2M_W\ll T\ll 2M_W/
\alpha_W$. However, the range of T may be taken to
 be $M_W\ll T\ll M_W/\alpha_W.$ [25]

We extrapolate this rate to get $\Gamma_R$, in a simple way, using
the following prescription:\\ (i) replace $M_W$ by $M_{WR}$,\\(ii)
write $T_{CR}= zM_{WR}(0)$,\\and (iii) include an overall factor b.

z is not known reliably, because, as yet, there isn't sufficient
experimental data to evaluate the full L-R Lagrangian, including the
Higg sector [27]. For large z, (34) shows that $M_{WR}
\approx M_{WR}(0)$. If z is too small, $M_{WR}$ will become 
imaginary. We will vary z between 2 and 10. The numerical work will
show that below z=2, the mass is not real, while there is little
change above z=10.

Whereas the exponential part in (33) will almost certainly be right
for $\Gamma_R$ (apart from the order one quantity $\bar{E}$), the
prefactor is bound to require considerable modification.  Considering
the prefactor to be a slowly varying quantity, whose main function is
to set the numerical scale of the essentially exponential variation
of $\Gamma_R$ with $(1/T)$, we will allow b to vary from
$10^{-3}\,-\,10^3$, i.e. the decoupling will be investigated with
anomalous rates for  right-handed neutrino reduction varying over 6
orders of magnitude around the rate obtained by simple substitution
of the right-handed W boson mass in the formula for the left-handed
case.

We have, then,\begin{eqnarray*}\lefteqn{
\Gamma_R=
\frac{\displaystyle (b\,1.4\times10^6)\,
M_{WR}(0)^7}{\displaystyle g^6\,T^3}[1- (\frac{\displaystyle
T}{\displaystyle zM_{WR}(0) })^2]^\frac{7}{2}}\\ & &  exp[-\frac
{\displaystyle 16\pi M_{WR}(0)}{\displaystyle
g^2T}\{1-(\frac{\displaystyle T}{\displaystyle
zM_{WR}(0)})^2\}^\frac{1}{2}]\end{eqnarray*} Introducing x and Y in
the above expression, the Boltzmann equation (25) becomes\[
\frac{\displaystyle dY} {\displaystyle
dx}=-f(x)(Y^2-Y_{eq}^2)-g(x)(
\frac{\displaystyle Y-Y_{eq}}{\displaystyle Y+Y_{eq}
}).\quad(35)\] \[ \mbox{From}\;(10), \; f(x)=\frac{\displaystyle
3.16\times 10^{17}}{
\displaystyle aM_{WR}(0)x^3}\qquad . \]  
\[ \Gamma_R\: \mbox{gives}\: g(x)=\frac{
\displaystyle b\,1.53\times 10^{23}x^7}{\displaystyle a^8M_{WR}(0)}
\{1-(\frac{\displaystyle a}
{\displaystyle zx})^2\}^\frac{7}{2}  exp[-
\frac{\displaystyle 118.98\,x}{\displaystyle a}
\{1-(\frac{\displaystyle a}{\displaystyle zx}
)^2\}^\frac{1}{2}],\] where $a=M/M_{WR}(0)$.

For $x<x_d$, this equation simplifies, as in Section II.C., to\[
\Delta=-\frac{\displaystyle Y_{eq}'} {\displaystyle
f(x)(2Y_{eq}+\Delta)+\frac{g(x)}{2Y_{eq} +\Delta}}.\qquad\quad(36)\]
We choose, again, as an approximate criterion for decoupling :\[
\Delta(x_d)\approx Y_{eq}(x_d)\quad
\Longrightarrow\quad Y(x_d)\approx 2Y_{eq}
(x_d).\] At decoupling, $Y_{eq}'=-Y_{eq}$. (36), then, leads to the
decoupling condition
\[3f(x_d)Y_{eq}(x_d)+\frac{\displaystyle g(x_d)}
{\displaystyle 3Y_{eq}(x_d)}=1.\quad(37)\]

Again, using the approximation\\ $Y_{\infty}\approx Y(x_d)\approx
2Y_{eq}(x_d)$,
\\the cosmological bound becomes, from (32),
\[1.62332\times10^6\,x_d^{\frac{3}{2}}
e^{-x_d}\,a\,M_{WR}(0)<1.\qquad\] At the bound,
\[1.62332\times10^6\,x_d^{\frac{3}{2}}
e^{-x_d}\,a\,M_{WR}(0)=1.\qquad(38)\]

First, we check the effect of z. Taking $M_{WR}(0)= 4000$ Gev and
b=1, we solve (37) and (38), numerically, to obtain values of $x_d=X$
and "a"=A, for which $
\Omega_Nh^2$ is just equal to 1. The results, displayed 
in Table II, show that, as z varies in the range $2\le z\le 10$, X
varies from 31.5 to 31.8, and A from 41 to 54. For z=1, $M_{WR}$ is
no longer real.  Also, as expected, z=100 gives for X and A
practically the same values as given by z=10.\begin{table}
\begin{center}{\bf TableII\\Effect
of uncertainty in $T_{CR}$}\\[1ex]\begin{tabular}[]{ccc}z&X&A
\\2&31.5&41\\3&31.6&48\\4&31.7&50
\\5&31.7&52\\6&31.7&53\\8&31.8&54
\\10&31.8&54\\100&31.8&55\end{tabular}
\end{center}\end{table}

Having seen that the effect of varying z is small, we set z=4 for
subsequent numerical work.

We next check that the bound obtained is actually a \it lower \rm
bound. We do this by varying "a" around the value A.  For each
assigned value of "a", we solve (37) for $x_d$, and evaluate
$\Omega_Nh^2$ for this $x_d$ from the LHS of (38).  The results, 
displayed in
Table III, show  that as "a" increases through the value A,
$\Omega_Nh^2$ falls through 1, from higher to lower
values.\begin{table}
\begin{center}{\bf Table III\\Mass Bound (with anomalous effects)}
\\[1ex]\begin{tabular}[]{ccc}
a& $x_d$ & $ \Omega_Nh^2$ \\100&234.8& $2.5\times 10^{-87}$
\\75&72.5& $9.8
\times 10^{-18}$ \\60&43.4& $1.6\times 
10^{-5}$ \\50.46&31.7&1.00\\40&22.1& $6.8\times 10^3$ \\25&19.6&
$4.3\times 10^4$ \\10&20.4& $8.2\times 10^3$
\end{tabular}\end{center}\end{table}

Finally, we vary b from $10^{-3}$ to $10^3$. The results are shown in
Table IV. We find that X changes from 31.8 to 31.6, and A changes
from 56 to 46. In every case, we have verified the nature of the
bound, numerically.  The results (not exhibited) parallel Table III.
The bound remains a \it lower \rm one. If, of course, smaller and
smaller values of b are considered, eventually the anomalous effects
term will be swamped by the second term in (25) or (28), and the
bound will revert to an upper one. However, the numerical work shows
that this does not happen even for b=$10^{-3}$.

\begin{table}\begin{center}{\bf Table IV\\Effect 
of overall uncertainty factor}\\[1ex]\begin{tabular}[]{ccc}b&X&A
\\0.001&31.8&56\\0.01&31.8&54\\0.1&31.7&52
\\1&31.7&50\\10&31.7&49\\100&31.6&47
\\1000&31.6&46\end{tabular}\end{center}\end{table}

It is necessary to verify that the restriction
$M_{WR}<T<M_{WR}/\alpha_W$ is satisfied. For the lower limit, the
worst case occurs when $M_{WR}
\approx M_{WR}(0)=4000$ Gev. Now, $T=4000\,A/X$, 
and the restriction is satisfied if $A>X$. A perusal of Tables III
and IV will show that this is, indeed, so, for the parameter ranges
considered by us. The stronger restriction, with $M_{WR}$ replaced by
$2M_{WR}$, is, however, not obeyed.

For the upper limit, the worst case occurs when z, and, hence,
$M_{WR}$ is the least. Taking z=2, X=31.5, and A=41, from Table II,
we find that $M_{WR}/\alpha_W\approx 90,000$ Gev, while $T\approx
5200$ Gev. The restriction is obeyed.

We check the kinematical constraint $E_{sp}>M$.  As
$E_{sp}=2(M_{WR}/\alpha_W)\bar{E}$, we look only at the case when
$M_{WR}$ is the least, viz.  z=2.  $E_{sp}$ comes out to be $>$
360,000 Gev, in this case, while, even for A=55, M=220,000 Gev, less
than $E_{sp}$, as required.
\bigskip

\begin{center}{\bf V. CONCLUSIONS}\end{center}

Analysing the decoupling of right-handed neutrinos with mass greater
than right-handed gauge boson masses, using normal electroweak
annihilation channels, we find that the cosmological bound
$\Omega_Nh^2<1$ leads to an {\it upper} bound on the right-handed
neutrino mass M, of about 700 Gev. What this really means is that a
right-handed neutrino mass greater than right-handed gauge boson
masses is unlikely to be allowed cosmologically.

If we now assume that anomalous B+L$-$violating processes work at the
right-handed symmetry-breaking scale by thermal diffusion over a
barrier, separating states of different B+L, in the same way as this
happens at the left-handed symmetry-breaking scale, then, we find
that it is possible to have a {\it lower} bound for a right-handed
neutrino mass greater than right-handed gauge boson masses.

A numerical extrapolation of the anomalous rate from the lower to the
higher energy scale, allowing a leeway of six orders of magnitude,
confirms this possibility. Taking $M_{WR}= 4\,$Tev, a lower bound
appears for the right-handed neutrino mass at about 50 times the
$W_R$ boson mass. However, in the absence of an explicit calculation
of the anomalous rate for the right-handed case,  the numbers must
only be considered as giving qualitative support to the idea that, at
Tev energy scales, anomalous generation plays an important part in
decoupling, and may take away cosmological obstacles to the existence
of right-handed neutrinos with mass greater than right-handed gauge
boson masses. To obtain reliable bounds, it is necessary to solve the
problem of constructing explicitly the sphaleron solution for the
right-handed case.\bigskip

\begin{center}{\bf ACKNOWLEDGEMENTS}\end{center}

The authors are grateful to Samir N. Mallik and Palash B. Pal of Saha
Institute of Nuclear Physics for a discussion of the problem at the
outset.  DRC wishes to thank A.Raychaudhuri of Calcutta University
and Avijit Lahiri of Vidyasagar College for helpful discussions, and
the former also for making available unpublished material. PA wishes
to thank Siddhartha Bhaumik and Kanad Ray of Presidency College,
Calcutta, for help in computing and preparation of the manuscript.
  

\bigskip

\begin{thebibliography}{99}
\bibitem{1}B.W. Lee and S. Weinberg, Phys. Rev. Lett. {\bf
 39}, 165 (1977).\bibitem{2}P. Hut, Phys. Lett.  {\bf 69B}, 85
(1977); K. Sato and M. Kobayashi, Progr. Theor. Phys.  {\bf 58}, 1775
(1977); M.I.  Vysotskii, A.D. Dolgov, and Ya-B. Zel'dovich, JETP
Lett. {\bf 26}, 188 (1977); D.A. Dicus, E.W.  Kolb, and V.L. Teplitz,
Phys. Rev. Lett.  {\bf 39}, 169 (1977).\bibitem{3}E.W. Kolb and K.A.
Olive, Phys. Rev. D {\bf 33}, 1202 (1986).
\bibitem{4}J.C. Pati and A. Salam, Phys.  Rev. D
{\bf 10}, 275 (1975); R.N.Mohapatra and J.C.Pati, {\it ibid} {\bf
11}, 566 (1975); {\it ibid}, {\bf 11}, 2558 (1975).\bibitem{5}G.
Senjanovic and R.N.  Mohapatra, Phys. Rev. D {\bf 12}, 1502
(1975).\bibitem {6}R.N. Mohapatra and G. Senjanovic, Phys. Rev. Lett.
{\bf 44}, 912 (1980); Phys. Rev. D {\bf 23}, 165 (1981).
\bibitem{7}M. Lindner and M. Weiser, Phys. Lett. B 
{\bf 383}, 405 (1996); R.R. Volkas, Phys. Rev. D {\bf 53}, 2681
(1996).\bibitem{8}J. Choi and R.R.  Volkas, Phys. Rev. D {\bf 48}
1258 (1993).
\bibitem{9} J.M. Fr$\acute{e}$re, L. Houart, J.M.
Moreno, J. Orloff, and M. Tytgat, Phys. Lett. B {\bf 314}, 289
(1993).
\bibitem{10}G. 't Hooft, Phys. Rev. Lett. {\bf 37}, 
8 (1976); Phys. Rev. D {\bf 14}, 3432 (1976); J.S. Bell and R.
Jackiw, Nuovo Cimento {\bf 51}, 47 (1969); S.L. Adler, Phys. Rev.
{\bf 177}, 2426 (1969); K. Fujikawa, Phys. Rev. Lett. {\bf 42}, 1195
(1979).\bibitem{11} V.A. Kuzmin, V.A.  Rubakov, and M.E.
Shaposhnikov, Phys.  Lett. {\bf 155B}, 36 (1985). For reviews and
discussion, see M. Shaposhnikov, Phys. Scr. T {\bf 36}, 1 (1991);
A.D. Dolgov, Phy. Rep. {\bf 222}, 309 (1992); {\it Baryon Number
Violation at the Electroweak Scale}, edited by Lawrence M. Krauss,
S-J Rey (World Scientific,1992); also ref.  [22].\bibitem{12}Palash
B. Pal and Rabindra N.  Mohapatra,{\it Massive Neutrinos in Physics
and Astrophysics}, 2nd ed., Chapter 6, p.100-109, Chapter 11,
p.196-199 (World Scientific, 1998).\bibitem{13}Kerson Huang, {\it
Statistical Mechanics}, 2nd ed. (John Wiley and Sons,
1987).\bibitem{14}Edward W. Kolb, Michael S.  Turner, {\it The Early
Universe}, Chapter 5, p.115-130 (Addison-Wesley Publishing Co.,
1990).\bibitem{15}S.M.  Bilenky and S.T. Petcov, Rev. Mod. Phys. {\bf
59}, 671 (1987); Ta-Pei Cheng and Ling-Fong Li, {\it Gauge Theory of
Elementary Particle Physics}, p.412-420 (Oxford University Press,
1991).\bibitem{16}R. Dashen, B.  Hasslacher, and A. Neveu, Phys. Rev.
D {\bf 10}, 4138 (1974); C.G. Callan, Jr., R. Dashen, and D.J. Gross,
{\it ibid} {\bf 17}, 2717 (1978); J.  Ambjorn, J.  Greensite, and C.
Petersen, Nucl.  Phys. {\bf B221}, 381 (1983); J. Kiskis, Phys.  Rev.
D {\bf 18}, 3690 (1978); N.H. Christ, {\it ibid} {\bf 21}, 1591
(1980).
\bibitem{17}N. Manton, Phys. Rev. D {\bf 28}, 2019 
(1983); F. Klinkhamer and N. Manton, {\it ibid} {\bf 30}, 2212
(1984).\bibitem{18}R.N. Mohapatra and X. Zhang, Phys. Rev. D {\bf
46}, 5331 (1992).\bibitem{19}J.  Langer, Ann. Phys. {\bf 54}, 258
(1969); Physica {\bf 73}, 6 (1974); I.  Affleck, Phys. Rev. Lett.
{\bf 46}, 388 (1981); E. Mottola, Nucl. Phys. {\bf B203}, 581 (1982);
A. Linde, Nucl. Phys. {\bf B216}, 421 (1981).
\bibitem{20}P. Arnold and L. McLerran, Phys. Rev. D
 {\bf 36}, 581 (1987); {\it ibid} {\bf 37}, 1020 (1988).
\bibitem{21}E.W. Kolb and M.S.Turner, Annu. Rev. Nucl. 
Part. Sci. {\bf 33}, p.667 (1983).\bibitem{22}A.G. Cohen, D.B.
Kaplan, and A.E. Nelson, Annu. Rev. Nucl. Part.  Sci. {\bf 43},
p.36-37 (1993).\bibitem{23}W. Fischler, G.F.  Giudice, R.G. Leigh,
and S. Paban, Phys. Lett. B {\bf 258}, 45 (1991); J. Harvey and M.
Turner, Phys.  Rev. D {\bf 42}, 3344 (1990); M. Fukugita and T.
Yanagida, Phys. Rev. D {\bf 42}, 1285 (1990); A.E. Nelson and
S.M.Barr, Phys. Lett. B {\bf 246}, 141 (1990).
\bibitem{24}L.Carson, Xu Li, L. McLerran, and R.T. Wang, 
Phys. Rev. D {\bf 42}, 2127 (1990).\bibitem{25}L.McLerran, Lectures
presented at {\it ICTP School on High Energy Physics}, July 1992,
Trieste, Italy.\bibitem{26}M. Dine, O.  Lechtenfeld, W. Fischler, and
J. Polchinski, Nucl.  Phys. {\bf B342}, 381 (1990); S.Yu.  Khlebnikov
and M.E.  Shaposhnikov, Nucl. Phys.  {\bf B308}, 885 (1988).
\bibitem{27}Gabriela Barenboim and Nuria Rius, Phys. Rev.
 D {\bf 58}, 065010 (1998).
\end{thebibliography}
\end{document}